\documentclass[aps,prl,a4paper,twocolumn,superscriptaddress,amsfonts,amsmath,amssymb,bm]{revtex4}

\begin{document}
\flushbottom 

\title{Elasticity-driven attraction between Abrikosov vortices in high-$\boldsymbol\kappa$ superconductors: Leading role of a non-core contribution}

\author{A. Cano}
\author{A. P. Levanyuk}
\affiliation{Departamento de F\'\i sica de la Materia Condensada, C-III, Universidad Aut\'onoma de Madrid, \\ E-28049 Madrid, Spain}
\author{S. A. Minyukov}
\affiliation{Institute of Crystallography, Russian Academy of Sciences,
Leninskii Prospect 59, Moscow 117333, Russia }

\date{ \today}

\begin{abstract}
We show that the strain-induced attraction between Abrikosov vortices has a non-core contribution overlooked up until now. This contribution is an example of the universal mechanism of soliton attraction in solids revealed in [Phys. Rev. B {\bf 66}, 14111 (2002)]. 
The resulting interaction energy is larger than that due to the vortex cores at least by a factor $\ln ^2 \kappa$. Consequently, the non-core contribution must be included, for instance, in discussions about the orientation of vortex lattices with respect to the crystal axes. It is shown to be also important when interpreting the thermal anomalies of the transition between superconducting and mixed states.
\end{abstract}

\pacs{PACS number(s): AA}

\maketitle

\section{I. Introduction}

\vspace{-.25cm}

Attractive interactions between Abrikosov vortices in type II superconductors have been a topic of great interest (see, e.g., Refs. \cite{Auer,Ullmaier,Kogan_Bulaevskii,Kogan,Blatter}). In particular, elasticity-driven attraction between vortices proves to have important experimental implications \cite{Ullmaier,Kogan_Bulaevskii,Kogan}. However, we shall show that this type of attraction was considerably underestimated until now. It is because it was assumed that vortex cores, considered as cylindrical inclusions of the normal phase, were the only sources of strain. In Ref. \cite{Levanyuk} a universal mechanism of strain-induced attraction between solitons has been revealed. This is operative even neglecting the strain induced by the soliton (vortex) cores (see below). In this Letter we show that this non-core mechanism gives the main contribution to the elasticity-driven attraction between Abrikosov vortices if the Ginzburg-Landau parameter $\kappa$ is large. 

We will consider elastically isotropic media only. Although this case is of some interest {\it per se}, its main convenience is that it represents a fairly simple case in which the leading role of the non-core contribution to the elasticity-driven attraction can be shown. With this result in mind it is natural to expect that in elastically anisotropic media, which have more practical interest, the same situation takes place (see Appendix).

We will discuss several consequences of this mechanism of attraction. First of all we will show that its contribution to the energy of the vortex lattice (VL) differs from that associated with the cores by a factor $\sim \ln^2 \kappa $ or even more. Previous authors \cite{Ullmaier,Kogan_Bulaevskii,Kogan} considered only the core contribution so they strongly underestimated this part of the VL energy of high-$\kappa$ superconductors. Despite this underestimation, the observed correlations between the VL and crystal lattice has been successfully explained by virtue of the strength of the strain-induced interaction in the considered materials. However, for other materials which can be studied in future it could be necessary to take into account both core and non-core contributions.

The existence of any attraction between vortices implies the discontinuity of the transition between superconducting and mixed states. As we will show, the discontinuity of the flux density associated with the strain-induced attraction between vortices is strong enough to be measurable in typical superconductors. There exist as well a latent heat associated with this discontinuity. It is expected to reveal itself experimentally as an anomaly in the specific heat. This anomaly should be taken into account when interpreting the experiments on the specific heat of mixed-state superconductors. 

The nature of strain-induced interactions between vortices can be easily illustrated considering a system with an infinite shear modulus ($\mu = \infty$) \cite{nota}. The only possible deformation of such a medium is a homogeneous dilatation $u$. The finite size of the sample is, in fact, taken into account by introducing this homogeneous deformation and considering a free crystal, i.e. introducing implicitly boundary conditions \cite{nota1}. Suppose that a density of vortices $n$ is created. The vortex self-energy depends, naturally, on $u$. Taking the state without vortices as the non-deformed one and omitting for a while the repulsive interaction between them, we present the change in the energy of the system per unit volume as $F (u) \simeq n(E_0+E_1 u) + Ku^2/2$, where $K$ is the bulk modulus and the terms in parentheses represent the above mentioned vortex self-energy. Minimizing with respect to $u$ one obtains the equilibrium deformation of the sample: $u_{\rm eq}=- n E _1/K$. Then, the change in energy becomes $F (u_{\rm eq}) = n E_0  - n^2 {E_1^2  /(2K)}$, where the second term represents the strain-induced attraction between vortices. Evidently one cannot forget that some vortex repulsion exists as well, a repulsion that provides a finite value of the equilibrium vortex density. Below we reproduce this result in more detail.

\vspace{-.25cm}
\section{II. Strain-induced attraction} 
\vspace{-.25cm}

We will start by showing that even neglecting the vortex cores, i.e. neglecting all sources of strain previously considered in Refs. \cite{Ullmaier,Kogan_Bulaevskii,Kogan}, there is a strain-induced attraction between vortices. Within Ginzburg-Landau theory the free energy density can be written as (see, e.g., Ref. \cite{Abrikosov})
\begin{align}
F= F_{GL} + F_{el}, 
\label{F_GL_el}\end{align}
where 
\begin{subequations}
\begin{gather}
F_{GL}= {1\over v}
 \int \Biggl[ a\left|\Psi \right|^2+{b\over 2}\left| \Psi\right|^4+{ \hbar ^2 \over 4 m} \left| \left( \nabla -{2ie\over \hbar c}{\mathbf A}\right)\Psi\right|^2 
\nonumber \\ \hspace{-3cm}
+{H^2 \over 8 \pi } \Biggr]dv,
\label{Iso_super}
\\ 
F_{el}= {1\over v}\int \left[ 
r u_{ll} \left| \Psi \right|^2
+\mu \left(u_{ik}-\frac 13\delta_{ik}u_{ll}\right)^2
+{K\over 2} u_{ll}^2
\right]dv.
\label{F_uik}
\end{gather}
\end{subequations}
Here $\Psi$ is the order parameter, $a=\alpha(T-T_c)$ is the only temperature dependent coefficient where $T_c$ is the critical temperature, and $v $ is the volume of the system. The elastic degrees of freedom are taken into account in Eq. \eqref{F_uik} where $u_{ik}$ is the strain tensor and summation over double indices is implied (see, e.g., Ref. \cite{Landau}). The first term in Eq. \eqref{F_uik} is responsible for the pressure dependence of the critical temperature and the change of the bulk modulus corresponding to the normal-superconductor phase transition.

The simplest way to work out the strain-induced vortex attraction is following the method given in Ref. \cite{Levanyuk}. To minimize the free energy over the elastic degrees of freedom we distinguish between homogeneous 
and inhomogeneous deformations \cite{Larkin_Pikin}:
\begin{align}
u_{ij}({\mathbf r}) &= u_{ij}^{(0)} +{i\over 2}
\sum_{{\mathbf k} \not = 0}\left[k_iu_j({\mathbf k}) 
+ k_ju_i({\mathbf k}) \right]e^{i{\mathbf k}\cdot {\mathbf r}}.
\label{}
\end{align}
Here, $u_{ij}^{(0)}$ represents the tensor of homogeneous deformations and $u_i (\mathbf k)$ the components of the displacement vector in Fourier space. Minimization of Eq. \eqref{F_uik} with respect to all elastic degrees of freedom gives 
\begin{align}
F_{el}=
-{r^2\over 2 K_{4/3}}
\langle\left| \Psi \right|^4\rangle
-{r^2\over 2K}{4 \mu \over 3 K_{4/3} }
\langle \left| \Psi\right|^2\rangle^2,
\label{F_nloc}\end{align}
where $K_{4/3}=K+4\mu/3$ and $\langle \dots \rangle$ means volume average. The contribution of the first term reduces to a renormalization of the coefficient $b$ in Eq. \eqref{Iso_super}. This renormalization disappears in the limit $\mu \to \infty$. Due to the second term the free energy becomes a non-local functional. Note that this non-locality remains as long as the shear modulus does not vanish.

Further minimization of Eq. \eqref{F_nloc} with respect to $\Psi$ is not straightforward due to its non-locality. However, if $\mu =\infty $ there is another minimization procedure that avoids the treatment of non-local equations. We first consider this case and after that we return to $\mu \not =\infty $. 

As we have mentioned, in this case the only possible deformation is a homogeneous dilatation $u$. Therefore, the free energy \eqref{F_GL_el} can be written as
\begin{align}
F ={1\over v}\int &\left[a(u)\left|\Psi \right|^2+{b\over 2}\left| \Psi\right|^4 +{ \hbar ^2 \over 4 m} \left| \left( \nabla -{2ie\over \hbar c}{\mathbf A}\right)\Psi\right|^2 \right.
\nonumber \\ 
&\left.+{H^2 \over 8 \pi } +{K\over 2} u^2
\right]dv,
\label{}
\end{align}
where $a(u)=a+ru$ and $u$ is a variational parameter. Fixing for a while this parameter, i.e. considering momentaneously a clamped sample, the form of the equations of equilibrium reduces to that of the Ginzburg-Landau equations \cite{Abrikosov}. Solving them one obtains, in particular, the free energy density close to the transition between superconducting and mixed states in terms of the magnetic induction $B=4 \pi \mathcal B$. For triangular VL's in high-$\kappa$ superconductors ($\ln \kappa \gg 1$) it can be written as
\begin{align}
F=&F_s(u)+{K \over 2}u^2+ {\mathcal  B} \left[H_{c1}(u)-H\right]
\nonumber \\ 
&+ \nu {\mathcal B}_0^{3/4}(u){\mathcal  B}^{5/4}\exp\left[{-\sqrt{ {\mathcal  B}_0(u)/{\mathcal  B} }}\right] ,
\label{f(n)}\end{align}
where $\nu=3^{3/2}({\pi /2 })^{1/2}$. The first two terms, being $F_s(u)=-a^2(u)/(2b)$, represent the free energy density in the superconducting state at $H=0$. The third term is proportional to the vortex self-energy, where $H_{c1}(u)$ is the magnetic field at which this self-energy changes its sign. It is known that for high-$\kappa$ superconductors the vortex self-energy comes mainly from contributions of the non-core region \cite{Abrikosov}. So taking $H_{c1}(u)=\{\phi_0/[4\pi \lambda ^2(u)]\}\ln\kappa$, where $\phi_0$ is the flux quantum and $\lambda(u)=\{mc^2b/[8\pi e^2|a(u)|]\}^{1/2}$ is the penetration length of the magnetic field, we shall reveal effects associated with these non-core contributions. The last term represents the repulsive interaction between vortices that takes place at low flux densities. Here ${\mathcal  B}_0(u) = \phi_0/[2\pi\sqrt3 \lambda^2(u)]$ defines the reference flux density.

Let us now consider free samples. Then Eq. \eqref{f(n)} has to be minimized with respect to $u$. The equilibrium deformation in the superconducting state is $u_s=ar/(b \widetilde K)$, where $\widetilde K = K - r^2/b$. In the mixed state there is, in addition, a deformation $ u_m= u - u_s$ as a result of the creation of vortices. Since it is small close to the transition between the superconducting and the mixed states, only lowest order terms are relevant and the $u_m$-dependence of the repulsion term can be neglected in Eq. \eqref{f(n)}. Thus, minimizing Eq. \eqref{f(n)} with respect to $u_m$ we obtain
\begin{align}
F\simeq &\gamma F_s+ {\mathcal  B}\left(\gamma H_{c1}- H\right) 
+  \nu (\gamma {\mathcal  B}_0)^{3/4}{\mathcal  B}^{5/4}\nonumber \\
&\times\exp\left({-\sqrt{ \gamma {\mathcal  B}_0/{\mathcal  B} }}\right) -\delta {\mathcal  B}^{2},
\label{f(n,infty)}
\end{align}
where $\gamma=K/\widetilde K$ and $\delta =\pi[r^2/(\widetilde K b)]\kappa^{-2} \ln^2 \kappa$ for high-$\kappa$ superconductors, i.e. $\ln \kappa = 2 \kappa^2 H_{c1}/H_{c2}$ \cite{Abrikosov} (here and in what follows the values $H_{c1}$, $H_{c2}$, etc. are refereed to the non-deformed state, $u=0$, if it is not explicitly indicated). The last term of Eq. \eqref{f(n,infty)} represents the strain-induced attraction between vortices. Mention that it disappears if the shear modulus goes to zero, i.e. it is associated with the solid-state elasticity. Let us recall that evaluating the vortex self-energy the vortex core was neglected. 

The free energy \eqref{f(n,infty)} could be obtained, in principle, from Eq. \eqref{F_nloc} with its coefficients corresponding to $\mu=\infty$, i.e. $r^2/(2K_{4/3})=0$ and $4\mu /(3K_{4/3})=1$. Note that there is no essential difference between the functional form of Eq. \eqref{F_nloc} for infinite and finite $\mu$. So we conclude that the free energy density of any isotropic type II superconductor has the form of Eq. \eqref{f(n,infty)} with the renormalized constants
$b'=b-{r^2/K_{4/3}}$ and $(r^2/ K)' = (r^2/ K)[4\mu / (3K_{4/3})]$.

\vspace{-.25cm}
\section{III. Core contribution} 
\vspace{-.25cm}

We will deduce the core contribution to the strain-induced attraction for $\mu =\infty $ in order to compare it with the non-core one \cite{nota1}. Following Ref. \cite{Kogan_Bulaevskii}, we model the vortices as cylinders of radius $\xi$ (coherence length) of normal phase inside a superconducting medium. Let us first consider a clamped superconducting medium. To accommodate a normal cylinder inside this medium, the cylinder should be deformed because of the difference between specific volumes of normal and superconducting phases $V_{n,s}$. Such a deformation is simply $u_0= {(V_n - V_s)/V_s}$ (if $\mu = \infty$ only homogeneous deformations are possible). Let us now consider a free sample designating as $n$ the density of cylinders (vortices). The elastic part of the free energy density can be written as 
\begin{align}
F \simeq n \pi \xi^2{K \over 2} (u-u_0)^2 + {K \over 2}u^2, 
\label{core_model}\end{align}
where $u$ is the deformation of the sample as a whole and it has been taken into account that the bulk moduli of both normal and superconducting phases are approximately equal ($K$). Minimizing Eq. \eqref{core_model} with respect to $u$ we obtain the equilibrium deformation of the sample: $u_m \simeq n \pi \xi^2 u_0 $. Therefore, the equilibrium free energy results to be 
\begin{align}
F\simeq
n \pi \xi^2{K u_0^2\over 2}
-n^2 \pi^2 \xi^4 {K u_0^2\over 2}. 
\label{attract_core}\end{align}
The second term of Eq. \eqref{attract_core} represents the attraction between vortices due to the core-induced strain. Taking into account that $n = B/\phi_0$ and (see, e.g., Ref. \cite{Kogan_Bulaevskii})  
\begin{align}
u_0= {V_n - V_s\over V_s} \simeq 
{H_c^2(T=0)\over 4\pi K T_c}{\partial T_c\over \partial u},
\end{align}
where $H_c^2=4\pi a^2/b$ and $({\partial T_c/\partial u}) =r/\alpha$, this second term can be written as  
\begin{align} 
F_{\rm core}^{\rm attr} \simeq- \delta \left( {H_{c2}^2 \ln \kappa\over 2^3\kappa^4 H_{c1}^2}\right)^2{\mathcal B}^2.\label{}\end{align}
For high-$\kappa$ superconductors one has $2\kappa^2 H_{c1}/H_{c2}=\ln \kappa$. Thus, the ratio between the core contribution to the strain-induced interaction and the non-core one [see Eq. \eqref{f(n,infty)}] is 
\begin{align}
F_{\rm core}^{\rm attr} /F_{\rm non-core}^{\rm attr}  \simeq 1/\ln ^{2} \kappa. 
\label{ratio}\end{align}

This ratio can be calculated more consistently close to $T_c$. In this region vortex cores can be successfully described within the Ginzburg-Landau theory, i.e. no model is necessary here. One has $H_{c1}(u)=\{\phi_0/[4\pi\lambda^2(u)]\}(\ln \kappa + 0.08)$ in Eq. \eqref{f(n)} (see, e.g., Ref. \cite{Abrikosov}). Further minimization of Eq. \eqref{f(n)} gives a free energy of the form Eq. \eqref{f(n,infty)} where the resulting coefficient $\delta$ includes both core and non-core contributions. It is such that 
\begin{align}
F_{\rm core}^{\rm attr} /F_{\rm non-core}^{\rm attr}  = 6.4 \times 10^{-3}/\ln ^{2} \kappa. 
\label{ratio1}\end{align}
As we see, the model of Ref. \cite{Kogan_Bulaevskii} strongly overestimate core contribution in this region. In any case, the strain-induced attraction energy due to the vortex cores results to be at least one order of magnitude lower than the non-core one for high-$\kappa $ superconductors.

The latter also shows that the contribution to the attraction energy due to the core-induced strain can be understood as a correction of the non-core one. Both contributions depends on orientation of the VL with respect to the crystal axes if one takes into account some elastic anisotropy, being this dependence the same (see Appendix). The ratio between both contributions remains given by Eq. \eqref{ratio}. In Refs. \cite{Ullmaier,Kogan_Bulaevskii,Kogan} it was suggested that the above mentioned dependence of the strain-induced interaction can explain the correlations between the vortex and crystal lattices. Mention that in Refs. \cite{Kogan_Bulaevskii,Kogan} this was asserted carrying out a detailed comparison between the London and the core-induced elastic energies of the VL's in $\rm NbSe_2$. The experimentally observed VL in $\rm NbSe_2$ does not correspond to the lowest energy one if only the London energy is taken into account. But including both London and elastic contributions, the experimentally observed VL coincides with the lowest energy one. This coincidence is a luck because the most important contribution to the elastic energy of the VL's, i.e. the non-core one, was overlooked. It might not be the case in other compounds \cite{Kogan97}, in which the non-core contribution could be essential.  

\vspace{-.25cm}
\section{IV. Thermal anomalies} 
\vspace{-.25cm}

As a result of the attraction between vortices, the transition from the superconducting to the mixed state has a discontinuous character. The conditions of minimum and continuity of the free energy determines the jump of the magnetic induction and the transition magnetic field. The attraction does not affect substantially the transition magnetic field which remains to be close to $H_{c1}$. The jump of the magnetic induction $\Delta B$ is such that
\begin{gather}
{\Delta B /B_0} 
\simeq \ln^{-2} \left\{ [ r^2  /(\widetilde Kb) ]
 \kappa^{-2} \ln ^2 \kappa \right\},
\label{B_tran}\end{gather}
where $B_0=4\pi\mathcal B_0$. The ratio $r^2/(\widetilde Kb)$ is the relative change in the bulk modulus at the normal-superconductor phase transition. Typically its order of magnitude is $r^2/(\widetilde K b) = 10^{-3} \div 10^{-6}$ \cite{Kogan_Bulaevskii,Bhattacharya88}. Thus, taking into account that $\kappa ^{-1}\ln \kappa \simeq 10^{-1}\div 10^{-2}$ ($\kappa = 10 \div 10^{3}$), the jump in the magnetic induction is expected to be $\Delta B \simeq 10^{-2}\div 10^{-3} B_0$. These values are high enough to be experimentally appreciable. 

There is a latent heat associated with this discontinuous transition. From Eq. \eqref{f(n,infty)} it can be written as $Q \simeq T \Delta B({\partial H_{c1}/\partial T})$. The contribution due to the strain-induced attraction can be roughly estimated as $\sim 1 ~\rm mJ ~ mol^{-1}$ in $\rm NbSe_2$ and $\sim 0.1 ~\rm mJ ~ mol^{-1}$ in $\rm YBa_2Cu_3O_{7-\delta}$. Since this phase transition is normally smeared, this latent heat will manifest itself as a specific anomaly. Having in mind the order of magnitude of the observed specific heat ($\sim 10 ~\rm mJ ~ mol^{-1} ~K^{-1}$ in $\rm NbSe_2$ and $\sim 1 ~\rm mJ ~ mol^{-1} ~ K^{-1}$ in $\rm YBa_2Cu_3O_{7-\delta}$, see e.g. Refs. \cite{Sonier99,Wright99}) one might conclude that this latent heat should be observed. 

Let us mention that measurements of the field-dependent specific heat have been proposed as a way to distinguish between $s$-wave and $d$-wave superconductors. The specific heat has, in both cases, contributions of terms which are proportional to the quasiparticles density of states localized in the vortex cores. Assuming that the vortex density depends linearly on the magnetic field, these terms are expected to be $\propto H$ in $s$-wave superconductors \cite{Fetter69}. In $d$-wave ones a weaker $\propto \sqrt H$ field dependence is expected as a result of the quasiparticle delocalization \cite{Volovik93}. In consequence, $s$ and $d$-wave superconductors are hopped to be distinguishable by presence or absence of curvature in the specific heat dependence on the external field.

In Refs. \cite{Ramirez96,Sonier99} it was pointed out that, even in $s$-wave superconductors, interactions between vortices could induce this curvature: these interactions provoke a nonlinear dependence of the vortex density on the external field. Only the repulsive interaction was explicitly mentioned, but our results indicate that the attractive interaction could be important as well. Taking all interactions into account, the specific heat anomaly associated with the transition between the superconducting and mixed states can be obtained from Eq. \eqref{f(n,infty)} in a similar way to that shown in Ref. \cite{Maki65}. 

Indeed, the dependence of the specific heat on the magnetic field might be affected by the smeared latent heat of the superconducting-mixed state transition. This contribution can not be quantified exactly, but the orders of magnitude obtained above indicate that it could be a significant contribution. As we see, when distinguishing between $s$-wave and $d$-wave superconductivity by way of measurements of specific heat anomalies much care should be taken. 

\vspace{-.25cm}
\section{V. Conclusions} 
\vspace{-.25cm}

We have shown the leading role of the non-core contribution to the strain-induced attraction between Abrikosov vortices in high-$\kappa$ superconductors. In the elastically isotropic case studied, the strain-induced interaction is due to finite size effects what previous authors neglected. The importance of the non-core contribution is also expected in a general case. Thermal anomalies associated with the transition between superconducting and mixed states have been discussed. The resulting jump in the magnetic induction and the latent heat of this transition was found strong enough to be measurable. The later might be important when interpreting experimental data on the specific heat of mixed-state superconductors.

\vspace{-.25cm}
\section{acknowledgments}
\vspace{-.25cm}

We acknowledge L.S. Froufe for useful discussions. A.P.L. was supported from the ESF programme Vortex Matter in Superconductors at Extreme Scales and Conditions (VORTEX). S.A.M. was supported from the Russian Fund for Fundamental Research (Grant No. 00-02-17746). 


\appendix
\vspace{-.25cm}
\section*{Appendix}
\vspace{-.25cm}

A full calculation of the strain-induced interaction between vortices in elastically anisotropic media is quite complicated and, to the best of our knowledge, has never been performed. The usual method to estimate this interaction consists in: (i) to calculate the interaction between two vortices in an infinite medium and (ii) to sum it over all vortex pairs (see, e.g., Ref. \cite{Kogan_Bulaevskii}). In what follows we shall analyze the first step of such a estimations.

For elastically anisotropic crystals, instead of Eq. \eqref{F_uik} one has
\begin{align}
F_{el}={1\over v}\int\left( 
r_{ik}u_{ik}|\Psi|^2 + {1\over 2 }\lambda _{iklm}u_{ik}u_{lm} 
\right)dv. 
\label{F_uik_ani}\end{align}
The minimization with respect to the elastic degrees of freedom can be worked out as before ($\mu \not = \infty$).  

Let us consider two well separated vortices, one at $\boldsymbol \rho =0$ and another at $\boldsymbol \rho =\boldsymbol \rho _0$ ($\rho_0 \gg \lambda$), directed along the $z$-axis. The strain-induced interaction between them can be written as \cite{Kogan_Bulaevskii}
\begin{align}
F_{el}^{pair}(\boldsymbol \rho _0)\negthickspace=\negthickspace
-\negthickspace \int \negthickspace
{d^2{\mathbf k}\over (2\pi)^2}
e^{i\mathbf k \cdot \boldsymbol \rho _0}r_{\mu \alpha}r_{\mu \beta }G_{\mu \nu}k_\alpha k_\beta |f(\mathbf k)|^2,
\label{F_12}\end{align}
where $G_{\mu \nu}^{-1}=\lambda_{\alpha \mu \nu \beta}k_\alpha k_\beta$ (Greek subscripts acquire only $x$ and $y$ values), and $f(\mathbf k)$ is the Fourier transform of the difference between the value of the order parameter far from the vortices and the square of the order parameter amplitude associated with one vortex.   

In Refs. \cite{Ullmaier,Kogan,Kogan_Bulaevskii} it was only considered the vortex core using the above described model of normal cylinders:
\begin{align}
f_{core}(\mathbf k)= \pi \xi^2 \Psi_0^2,
\label{}\end{align}
where $\Psi_0^2 = |a|/b$. Let us neglect this contribution, focusing our attention in the non-core one. Then, because the order parameter amplitude is such that \cite{Abrikosov}
\begin{align}
f(\boldsymbol \rho)/\Psi_0^2\simeq \begin{cases}
1,& \lambda \ll \rho, \\
1-\left(\xi /\rho \right)^2,
& \xi \ll \rho \ll \lambda, \\
\rho/\xi,          & \rho \ll \xi,\\
\end{cases}
\label{f_rho}\end{align}
we can take
\begin{align}
f_{non-core}(\mathbf k)&\simeq \xi^2\Psi_0^2 
\int_\xi^\lambda \negthickspace \int_0^{2\pi}{e^{ik\rho\cos \theta}\over \rho}d\rho d \theta 
\nonumber \\&
= 2 \pi \xi^2\Psi_0^2 
\int_\xi^\lambda \negthickspace {J_{0}\left(k\rho\right)\over \rho}
d\rho.
\label{Bessel}\end{align}
Due to the separation between vortices, the most important contribution to Eq. \eqref{F_12} comes from $k\ll\lambda^{-1}$. For these values, the argument of the Bessel function in Eq. \eqref{Bessel} is $k\rho \ll 1 $ inside of the integration interval, so
\begin{align}
f_{non\negthickspace-core}(\mathbf k)&\simeq 2 \pi \xi^2\Psi_0^2 
\int_\xi^\lambda \negthickspace {J_{0}\left(0\right)\over \rho}
d\rho = 2 \pi \xi^2 \Psi_0^2  \ln \kappa. 
\label{}\end{align}
As we see, the non-core contribution to $F_{el}^{pair}$ prevails over the previously reported core one by virtue of the high value of the Ginzburg-Landau parameter $\kappa$.

Mention that close to $T_c$ the expression \eqref{f_rho} can also be used in order to estimate the core contribution, i.e. no model is necessary here. Thus one obtains
\begin{align}
f_{core}(\mathbf k)&\simeq {2 \pi \Psi_0^2 \over \xi } 
\int_0^\xi \negthickspace (\xi\rho - \rho^2)
d\rho = {\pi \over 3}\xi^2 \Psi_0^2. 
\label{}\end{align}
Therefore, the ratio between core and non-core contributions to $F_{el}^{pair}$ results to be $\sim 1/(2\ln ^{2}\kappa)$ (the model of normal cylinders slightly overestimates the core contribution in this region).


\begin{thebibliography}{999}
\bibitem{nota1} Image forces (finite size effects) was mentioned already in Ref. \cite{Ullmaier} but was not treated explicitly. In Refs. \cite{Kogan_Bulaevskii,Kogan} they was neglected at all.
\end{thebibliography}
\end{document}